\documentstyle[epsfig,12pt]{article}

\catcode`\@=11
\long\def\@makefntext#1{
\protect\noindent \hbox to 3.2pt {\hskip-.9pt
$^{{\ninerm\@thefnmark}}$\hfil}#1\hfill}		

\def\@makefnmark{\hbox to 0pt{$^{\@thefnmark}$\hss}}  

\def\ps@myheadings{\let\@mkboth\@gobbletwo
\def\@oddhead{\hbox{}
\rightmark\hfil\ninerm\thepage}
\def\@oddfoot{}\def\@evenhead{\ninerm\thepage\hfil
\leftmark\hbox{}}\def\@evenfoot{}
\def\sectionmark##1{}\def\subsectionmark##1{}}

\renewcommand{\thefootnote}{\fnsymbol{footnote}}

\newcounter{sectionc}\newcounter{subsectionc}\newcounter{subsubsectionc}
\renewcommand{\section}[1] {\vspace*{0.6cm}\addtocounter{sectionc}{1}
\setcounter{subsectionc}{0}\setcounter{subsubsectionc}{0}\noindent
	{\normalsize\bf\thesectionc. #1}\par\vspace*{0.4cm}}
\renewcommand{\subsection}[1] {\vspace*{0.6cm}\addtocounter{subsectionc}{1}
	\setcounter{subsubsectionc}{0}\noindent
	{\normalsize\it\thesectionc.\thesubsectionc. #1}\par\vspace*{0.4cm}}
\renewcommand{\subsubsection}[1]
{\vspace*{0.6cm}\addtocounter{subsubsectionc}{1}
	\noindent {\normalsize\rm\thesectionc.\thesubsectionc.\thesubsubsectionc.
	#1}\par\vspace*{0.4cm}}

\newcounter{appendixc}
\newcounter{subappendixc}[appendixc]
\newcounter{subsubappendixc}[subappendixc]

\renewcommand{\appendix}[1] {\vspace*{0.6cm}
        \refstepcounter{appendixc}
        \setcounter{figure}{0}
        \setcounter{table}{0}
        \setcounter{equation}{0}
        \renewcommand{\thefigure}{\Alph{appendixc}.\arabic{figure}}
        \renewcommand{\thetable}{\Alph{appendixc}.\arabic{table}}
        \renewcommand{\theappendixc}{\Alph{appendixc}}
        \renewcommand{\theequation}{\Alph{appendixc}.\arabic{equation}}
        \noindent{\bf Appendix \theappendixc #1}\par\vspace*{0.4cm}}

\def\abstracts#1{{

\centering{\begin{minipage}{12.2truecm}\footnotesize\baselineskip=12pt\noindent
	\centerline{\footnotesize ABSTRACT}\vspace*{0.3cm}
	\parindent=0pt #1
	\end{minipage}}\par}}

\renewenvironment{thebibliography}[1]
	{\begin{list}{\arabic{enumi}.}
	{\usecounter{enumi}\setlength{\parsep}{0pt}
\setlength{\leftmargin 1.25cm}{\rightmargin 0pt}
	 \setlength{\itemsep}{0pt} \settowidth
	{\labelwidth}{#1.}\sloppy}}{\end{list}}

\newcounter{itemlistc}
\newcounter{romanlistc}
\newcounter{alphlistc}
\newcounter{arabiclistc}

\newcommand{\tcaption}[1]{
        \refstepcounter{table}
        \setbox\@tempboxa = \hbox{\footnotesize Table~\thetable. #1}
        \ifdim \wd\@tempboxa > 6in
           {\begin{center}
        \parbox{6in}{\footnotesize\baselineskip=12pt Table~\thetable. #1}
            \end{center}}
        \else
             {\begin{center}
             {\footnotesize Table~\thetable. #1}
              \end{center}}
        \fi}

\def\@citex[#1]#2{\if@filesw\immediate\write\@auxout
	{\string\citation{#2}}\fi
\def\@citea{}\@cite{\@for\@citeb:=#2\do
	{\@citea\def\@citea{,}\@ifundefined
	{b@\@citeb}{{\bf ?}\@warning
	{Citation `\@citeb' on page \thepage \space undefined}}
	{\csname b@\@citeb\endcsname}}}{#1}}

\newif\if@cghi
\def\cite{\@cghitrue\@ifnextchar [{\@tempswatrue
	\@citex}{\@tempswafalse\@citex[]}}
\def\citelow{\@cghifalse\@ifnextchar [{\@tempswatrue
	\@citex}{\@tempswafalse\@citex[]}}
\def\@cite#1#2{{$\null^{#1}$\if@tempswa\typeout
	{IJCGA warning: optional citation argument
	ignored: `#2'} \fi}}
\newcommand{\citeup}{\cite}

 1
 1
 1

\font\ninerm=cmr9

\begin{document}

\newcommand{\st}{\scriptstyle}
\newcommand{\sst}{\scriptscriptstyle}
\newcommand{\mco}{\multicolumn}
\newcommand{\epp}{\epsilon^{\prime}}
\newcommand{\vep}{\varepsilon}
\newcommand{\ppg}{\pi^+\pi^-\gamma}
\newcommand{\vp}{{\bf p}}
\newcommand{\ko}{K^0}
\newcommand{\kb}{\bar{K^0}}
\newcommand{\al}{\alpha}
\newcommand{\ab}{\bar{\alpha}}
\def\CPbar{\hbox{{\rm CP}\hskip-1.80em{/}}}

\def\solid{(---------)}
\def\dashes{($-~-~-~-$)}
\def\dots{($\cdot~\cdot~\cdot~\cdot~\cdot~\cdot\,$)}
\def\daashdash{(-----~$-$~-----~$-$)}
\def\dotdash{($\cdot~-~\cdot~-$)}
\def\dotdotdash{($\cdot~\cdot~-~\cdot~\cdot~-$)}
\def\dotdotdotdash{($\cdot~\cdot~\cdot~-$)}
\def\dotdotdotdotdash{($\cdot~\cdot~\cdot~\cdot~-$)}
\def\dotdashdotdotdash{($\cdot~-~\cdot~\cdot~-$)}
\def\dotdotdotdashdash{($\cdot~\cdot~\cdot~-~-$)}
\def\dotdashdash{($\cdot~-~-~\cdot~-~-$)}
\def\dotdotdashdash{($\cdot~\cdot~-~-$)}
\def\ee#1{\times 10^{{#1}}}
\def\tev{~{\rm TeV}}
\def\pbi{~{\rm pb}^{-1}}
\def\fbi{~{\rm fb}^{-1}}
\def\fb{~{\rm fb}}
\def\pb{~{\rm pb}}
\def\hl{h^0}
\def\hh{H^0}
\def\ha{A^0}
\def\mhl{m_{\hl}}
\def\mhh{m_{\hh}}
\def\mha{m_{\ha}}
\def\hpm{H^{\pm}}
\def\mhpm{m_{\hpm}}
\def\dphi{\delta\phi}
\def\gam{\gamma}
\def\lplm{l^+l^-}
\def\sinb{\sin\beta}
\def\cosb{\cos\beta}
\def\rta{\rightarrow}
\def\mupmum{\mu^+\mu^-}
\def\taup{\tau^+}
\def\taum{\tau^-}
\def\mtau{m_\tau}
\def\nsdi{\nsd^1}
\def\nsdii{\nsd^2}
\def\nsdiii{\nsd^3}
\def\gaml{\kappa}
\def\sig{\sigma}
\def\fhalfs#1{F^s_{1/2}(#1)}
\def\fhalfp#1{F^p_{1/2}(#1)}
\def\fone#1{F_1(#1)}
\def\fzero#1{F_0(#1)}
\def\dlgamgam{dL_{\gam\gam}}
\def\lame{\lam_e}
\def\zo{\zeta_0}
\def\zi{\zeta_1}
\def\zii{\zeta_2}
\def\ziii{\zeta_3}
\def\zteta{\widetilde\zeta}
\def\zto{\zteta_0}
\def\zti{\zteta_1}
\def\ztii{\zteta_2}
\def\ztiii{\zteta_3}
\def\anti{\overline}
\def\hn{H}
\def\hnp{\hn^{\prime}}
\def\hnpp{\hn^{\prime\,\prime}}
\def\mhn{m_\hn}
\def\epsi{{\bf e}}
\def\epsii{{\bf \wtilde e}}
\def\lami{\lam}
\def\lamii{\wtilde\lam}
\def\deliii{\delta_{\lami\lamii}}
\def\mpp{M_{++}}
\def\mmm{M_{--}}
\def\mpm{M_{+-}}
\def\mmp{M_{-+}}
\def\absq#1{\left| #1 \right|^2}
\def\asym{{\cal A}}
\def\asymi{{\cal A}_1}
\def\asymii{{\cal A}_2}
\def\asymiii{{\cal A}_3}
\def\aimax{|\asymi|_{max}}
\def\aiimax{|\asymii|_{max}}
\def\aiiimin{|\symiii|_{min}}
\def\wm{W^-}
\def\wp{W^+}
\newcommand{\nc}{\newcommand}
\nc{\pbarn}{\hbox {pb}}
\nc{\lumun}{\;{\hbox {pb}^{-1}}{\hbox {yr}^{-1}}}
\nc{\hc}{\hbox {h.c.}}
\nc{\re}{\hbox {Re}}
\nc{\im}{\hbox {Im}}
\nc{\mev}{\hbox {MeV}}
\nc{\gev}{\;\hbox {GeV}}
\nc{\etal}{\hbox{et al.}}
\nc{\prdj}[1]{{ \it Phys.~Rev.}~{\bf D{#1}}}
\nc{\prlj}[1]{{ \it Phys.~Rev.~Lett.}~{\bf {#1}}}
\nc{\plbj}[1]{{ \it Phys.~Lett.}~{\bf {#1B}}}
\nc{\npbj}[1]{{ \it Nucl.~Phys.}~{\bf B{#1}}}
\nc{\ptpj}[1]{{ \it Prog.~Theor.~Phys.}~{\bf {#1}}}
\nc{\zpcj}[1]{{ \it Z.~Phys.}~{\bf C{#1}}}
\nc{\mplaj}[1]{{ \it Mod.~Phys.~Lett.}~{\bf {#1A}}}
\nc{\beq}{\begin{equation}}
\nc{\eeq}{\end{equation}}
\nc{\bea}{\begin{eqnarray}}
\nc{\eea}{\end{eqnarray}}
\nc{\baa}{\begin{array}}
\nc{\eaa}{\end{array}}
\nc{\ra} {\rightarrow}
\nc{\ttbar}{t\bar{t}}
\nc{\bbbar}{b\bar{b}}
\nc{\tanb} {\tan \beta}
\nc{\twbdec} {t\rightarrow W^+ b}
\nc{\tbwbdec} {\bar{t} \rightarrow W^- \bar{b}}
\nc{\hprod} {e^+e^- \ra Z^\ast \ra H Z}
\nc{\epem} {e^+e^-}
\nc{\wpwm} {W^+W^-}
\nc{\tbar} {\bar{t}}
\nc{\bbar} {\bar{b}}
\nc{\wpp} {W^+}
\nc{\mt}{m_t}
\nc{\mts}{m_t^2}
\nc{\mj}{m_j}
\nc{\mjs}{m_j^2}
\nc{\mi}{m_i}
\nc{\mis}{m_i^2}
\nc{\mw} {m_W}
\nc{\mws} {m_W^2}
\nc{\mz} {m_Z}
\nc{\mzs} {m_Z^2}
\nc{\mh} {m_H}
\nc{\mhs} {m_H^2}
\nc{\hdec}{H \ra t\bar{t}}
\nc{\ttbardec}{\ttbar \ra W^+W^-\bbbar}
\nc{\wwbb}{W^+W^-\bbbar}
\nc{\po}{\Phi_1}
\nc{\pot}{{\tilde{\Phi}}_1}
\nc{\pod}{\Phi_1^\dagger}
\nc{\pht}{\Phi_2}
\nc{\phtd}{\Phi_2^\dagger}
\nc{\phtt}{{\tilde{\Phi}}_2}
\nc{\popo}{\po^\dagger\po}
\nc{\phtpt}{\pht^\dagger\pht}
\nc{\popt}{\po^\dagger\pht}
\nc{\phtpo}{\pht^\dagger\po}
\nc{\as}{{\cal A}_{tt}}
\nc{\bt}{\beta_t}
\nc{\bts}{\beta_t^2}
\nc{\bz}{\beta_Z}
\nc{\bzs}{\beta_Z^2}
\nc{\bw}{\beta_W}
\nc{\bws}{\beta_W^2}
\nc{\sq}{\sqrt{2}}
\nc{\wphel}{\mbox{$\lambda_{W^+}$}}
\nc{\wmhel}{\mbox{$\lambda_{W^-}$}}
\nc{\bhel} {\mbox{$h_{b}$}}
\nc{\thel} {\mbox{$h_{t}$}}
\nc{\nsd} {N_{SD}}
\nc{\ntt} {N_{tt}}
\nc{\gsim} {>_{\!\!\!\sim}}
\nc{\lsim} {<_{\!\!\!\sim}}
\nc{\ie}{{\it i.e.}}
\nc{\eg}{{\it e.g.}}

\centerline{\normalsize\bf CP VIOLATION IN FERMIONIC DECAYS}
\baselineskip=22pt
\centerline{\normalsize\bf OF HIGGS BOSONS\footnote{Presented by
B. Grz\c{a}dkowski}}
\baselineskip=22pt
\centerline{\footnotesize B. Grz\c{a}dkowski}
\baselineskip=13pt
\centerline {\footnotesize\it Institute of Theoretical Physics}
\centerline{\footnotesize\it Department of Physics, Warsaw University,
Warsaw, PL-00-681 Poland }
\baselineskip=12pt
\centerline{\footnotesize E-mail: bohdang@fuw.edu.pl}
\vspace*{0.1cm}
\centerline{\footnotesize and}
\vspace*{0.1cm}
\centerline{\footnotesize J.F. Gunion}
\baselineskip=13pt
\centerline{\footnotesize\it Davis Institute for High Energy Physics}
\centerline{\footnotesize\it Department of Physics,
University of California, Davis, CA 95616 U.S.A.}
\centerline{\footnotesize E-mail: jfgucd@ucdhep.ucdavis.edu}
\vspace*{0.5cm}
\abstracts{We demonstrate that decay angle correlations in $\taum\taup$
and $t\anti t$ decay modes could allow a determination of whether or
not a neutral Higgs boson is a CP eigenstate. Sensitivity of
the correlations is illustrated in the case of the
$\epem\rta Z \hn$ and $\mupmum\rta \hn$ production processes
for a two-doublet Higgs model with CP-violating neutral sector.
A very useful technique for minimizing `depolarization' factor
suppressions of the correlations in the $t\anti t$ mode is introduced.}
\normalsize\baselineskip=15pt
\setcounter{footnote}{0}
\renewcommand{\thefootnote}{\alph{footnote}}
\vspace*{.5cm}
Determination of the CP nature of any neutral Higgs boson that is
directly observed will be crucial to fully unraveling the nature of the Higgs
sector. It is entirely
possible to have either explicit or spontaneous CP violation in the neutral
Higgs sector. Indeed, the simplest non-supersymmetric two-Higgs-doublet
model (2HDM) and the supersymmetric Higgs two-doublet plus singlet model
both allow for Higgs mass eigenstates of impure CP nature\cite{hhg}.
Here we shall focus
on the 2HDM, in which CP violation results in
three neutral states, $\hn_{i=1,2,3}$, of mixed CP character.

The most direct probe of CP violation is provided by
comparing the Higgs boson production rate
in collisions of two back-scattered-laser-beam
photons of various different polarizations\cite{bgbslaser}.
A certain difference in rates for different photon helicity choices
is non-zero only if CP violation is present, and has a
good chance of being of measurable size for many 2HDM parameter choices.

Correlations between decay products can also probe the
CP nature of a Higgs boson. In this paper we focus
on effects that arise entirely at tree-level. For the dominant
two-body decays of a Higgs boson, one can define
appropriate observables if we are able to
determine the rest frame of the Higgs boson and if the secondary
decays of the primary final state particles allow
an analysis of their spin or helicity directions. An obvious example
is to employ correlations between the decay planes of the
decay products of $WW$ or $ZZ$ vector boson pairs and/or energy
correlations among the decay
products\cite{nelson,duncan,ck,ckp,soni,bargeretal,skjold}.
However, these will not be useful for a purely CP-odd $\hn$ (which
has zero tree-level $WW,ZZ$ coupling and thus decays primarily to $F\anti F$)
or for a mixed-CP $\hn$ in the (most probable) case where the CP-even component
accounts for essentially all of the $WW,ZZ$ coupling strength (thereby yielding
`apparently CP-even' correlations).
In contrast, $\hn$ decays to $\taum\taup$
or $t\anti t$, followed by $\tau$ or $t$ decays, do, in principle,
allow equal sensitivity to the CP-even and CP-odd components of
a given Higgs boson.

Indeed, a $\hn$ eigenstate couples to $F\anti F$ according to:
$ {\cal L}\propto \anti F (a+ib\gamma_5) F \hn\,,$
which yields
\beq
\langle F_+ \anti F_+ | \hn \rangle\propto b+ia\beta_F;\quad
\langle F_- \anti F_- | \hn \rangle\propto b-ia\beta_F\,,
\label{amps}
\eeq
where $\beta_F=\sqrt{1-4m_F^2/\mhn^2}$, helicity-flip amplitudes
being zero. The crucial point is that, in general,
$a$ and $b$ are of comparable magnitude
in a CP-violating 2HDM.

Here, we shall
present a unified treatment aimed at realistically evaluating
the possibility of using correlations in $\hn\rta \taum\taup$ and $t\anti t$
final states to determine if
a decaying Higgs boson is a mixed CP eigenstate, thereby directly
probing for the presence of CP violation in the Higgs sector.

An efficient framework for our analysis is that developed
in Refs.~\citeup{kw,kksz}. Consider the charged current decay $F\rta R f$,
where $F$
is a heavy fermion, $f$ is a light fermion whose mass can be neglected,
and $R$ can be either a single particle or a multiparticle state with known
quantum numbers and, therefore, calculable coupling to the charged weak
current.
(Examples are $\tau\rta R \nu$, where $R=\pi,\rho,A_1,\ldots$, and $t\rta W b$,
where $W$ decays to a fermion plus anti-fermion.)  For the $R$'s of interest,
the form of the hadronic current $J_\mu$, deriving from the standard
$V-A$ interaction for the
$J_\mu\equiv\langle R\vert V_\mu - A_\mu \vert 0 \rangle$ coupling,
is completely determined in terms of the final particle momenta.
Using the particle symbol to denote also its momentum, and defining
\beq
\Pi_\mu=4{\rm Re}\,J_\mu f\cdot J^*-2 f_\mu J\cdot J^*,\quad
\Pi_\mu^5=2\epsilon_{\mu\rho\nu\sigma} {\rm Im}\,J^\rho J^{*\,\nu} f^\sigma,
\label{pidefs}
\eeq
all useful correlations in $\hn\rta F\anti F$ decay can
be obtained by employing the quantities
\beq
\omega=F\cdot(\Pi-\Pi^5),\quad R_\mu=m_F^2(\Pi-\Pi^5)_\mu - F_\mu F\cdot
(\Pi-\Pi^5)\,,
\label{omegardefs}
\eeq
and their $\anti F$ analogues.
In the $F$ rest frame, $R_0=0$, $\vec{R}=m_F^2(\vec\Pi-\vec\Pi^5)$, and
$|\vec{R}|=m_F\omega$. In fact,
$\vec S_F=\vec{R}/(m_F\omega)$ acts as an effective spin direction
($|\vec S_F|^2=1$) when in the $F$ rest frame.

Let us give some illustrative examples. For $\taum\rta \pi^-\nu$
decay, $J_\mu\propto \pi_\mu^-$
and $\vec S_{\taum}=\hat{\pi^-}$ is the unit vector pointing
in the direction of the $\pi^-$'s three momentum (using
angles defined in the $\taum$ rest frame). For $\taum\rta \rho^-\nu\rta
\pi^-\pi^0\nu$, $J_\mu\propto (\pi^--\pi^0)_\mu$, yielding
$\Pi_\mu\propto 4(\pi^--\pi^0)_\mu \nu\cdot(\pi^--\pi^0)+2\nu_\mu m_\rho^2$,
and, thence,
$\vec S_F\propto m_\tau(\vec{\pi^-}-\vec{\pi^0})(E_{\pi^-}-E_{\pi^0})
+  \vec{\nu}m_\rho^2/2\,,$
where the pion energies and directions are defined in the $\taum$ rest frame.
For $t\rta W^+ b\rta l^+\nu b$,
$J_\mu\propto \anti u(\nu)\gamma_\mu (1-\gamma_5)v(l^+)$,
and $\Pi_\mu\propto l^+_\mu \nu\cdot b + \nu_\mu l^+\cdot b$,
$\Pi_\mu^5\propto \nu_\mu l^+\cdot b-l^+_\mu \nu\cdot b $, so that
$\Pi_\mu-\Pi_\mu^5\propto l^+_\mu$, implying $\vec{S_t}=\hat l^+$
in the $t$ rest frame.

If the full $(\Pi-\Pi^5)_\mu$ can be determined on an event-by-event
basis, then we can define
the `effective spin' vectors $\vec S_F$ and $\vec S_{\anti F}$
{\it for each event}, and the distribution of the Higgs decay products takes
the very general form
\bea
dN & \propto & \Bigl[(b^2+a^2\beta_F^2)(1+\cos\theta\cos\anti\theta)
        +(b^2-a^2\beta_F^2)\sin\theta\sin\anti\theta\cos(\phi-\anti\phi)
        \nonumber\\
&&-2ab\beta_F\sin\theta\sin\anti\theta\sin(\phi-\anti\phi)
        \Bigr]d\cos\theta d\cos\anti\theta d\phi d\anti\phi\,,
\label{dnform}
\eea
where $\theta,\phi$ and $\anti\theta,\anti\phi$ define the angles
of $\vec S_F$ and $\vec S_{\anti F}$ in the $F$ and $\anti F$ rest frames,
respectively,
{\it employing the direction of $F$ in the $\hn$ rest frame as the
coordinate-system-defining $z$ axis}.

If we cannot determine $(\Pi-\Pi^5)_\mu$ for each event, then Eq.~\ref{dnform}
must be modified.  An extreme example is $F\rta R f$ decay where the $R$
decay products are not examined.  In this case
the angles of $R$ in the $F$ rest frame would be employed in Eq.~\ref{dnform},
and `depolarization' factors arise as a result of event averaging.
In deriving Eq.~\ref{dnform}, the angular independent term is actually
multiplied by $(m_F\omega_F)(m_F\omega_{\anti F})$ and the
$\cos\theta\cos\anti\theta$, $\sin\theta\sin\anti\theta
\sin(\phi-\anti\phi)$ and
$\sin\theta\sin\anti\theta\cos(\phi-\anti\phi)$
terms by $|\vec R_F||\vec R_{\anti F}|$.
On an event-by-event basis the ratio of these coefficients is unity,
as outlined earlier.  When averaged over events, this is no longer true.
Consequently, when event averaging (denoted by $\langle\ldots\rangle$)
all the angle-dependent terms in Eq.~\ref{dnform} must be multiplied by
$D_F\equiv\langle |\vec R_F|\rangle/(m_F\langle \omega_F\rangle)$ and/or
its $D_{\anti F}$ analogue, relative to the angle-independent term.
We define $D\equiv D_F D_{\anti F}$.

At first sight, the necessity of event averaging arises
in the case of the $t\anti t$ final state, for which we will find that
we must have one top decay leptonically and the other hadronically
in order to define the $t\anti t$ line of flight
and, thereby, appropriate angles in Eq.~\ref{dnform}.
For the hadronically decaying top, the problem is to distinguish
the quark vs. anti-quark jet coming from the $W$ so as to construct
$(\Pi-\Pi^5)_\mu$ (which is proportional to the $\wp$ ($\wm$)
anti-quark (quark)
momentum for $t$ ($\anti t$) decay) for each event.
If we simply sum over all $W$ decay
product configurations, then the appropriate depolarization factor
is easily computed by using $J_\mu\propto \epsilon^W_\mu$ and
summing over $W$ polarizations.
One finds $D_t=(m_t^2-2\mw^2)/(m_t^2+2\mw^2)\sim 0.4$
for $\mt=174\gev$.
Similarly, for $\tau\rta R\nu$ where $R$ is spin-1.

Let us now specify our procedure for isolating the coefficients of the
$\cos(\phi-\anti\phi)$ and $\sin(\phi-\anti\phi)$ angular correlation
terms. Defining $c\equiv \cos\theta$, $\anti c\equiv\cos\anti\theta$,
$s\equiv \sin\theta$, $\anti s\equiv\sin\anti\theta$,
$c_{\phi}\equiv\cos\delta\phi$, $s_\phi\equiv \sin\delta\phi$
(where $\delta\phi\equiv \anti\phi-\phi$), and $d\Omega\equiv dc\, d\anti c \,
d\delta\phi$, and including a possible depolarization factor, we have
\beq
{1\over N}{dN\over d\Omega}={1\over 8\pi} \left[ 1+ D c\anti c
+ \rho_1 s \anti s s_\phi+\rho_2 s\anti s c_\phi\right]\,,
\label{phidist}
\eeq
where
\beq
\rho_1\equiv D{2ab\beta_F\over (b^2+a^2\beta_F^2)}\,,\quad
\rho_2\equiv D{(b^2-a^2\beta_F^2)\over (b^2+a^2\beta_F^2)}\,.
\label{rhodefs}
\eeq
For a CP-conserving Higgs sector, either $a=0$ or $b=0$ implying
$\rho_1=0$ and $|\rho_2|=D$. For a CP-mixed eigenstate,
both $a$ and $b$ are non-zero. Thus $\rho_1\neq 0$ provides
an unequivocable signature for CP violation in the Higgs sector, while
the difference $D-|\rho_2|$  also provides a measure
of Higgs sector CP violation. (Indeed, $\rho_1$ and $\rho_2$ are not
independent; $\rho_1^2+\rho_2^2=D^2$.) Values of $\rho_1\sim D$
and $\rho_2\sim 0$ are common in an unconstrained 2HDM.

To isolate $\rho_1$ and $\rho_2$, we define projection functions
$f_{1,2}(\theta,\anti\theta,\delta\phi)$ such that
$\int f_{1,2}d\Omega=0$, $\int f_{1,2} c\anti c d\Omega=0$,
$\int f_{1} s \anti s s_\phi d\Omega=8\pi$,
$\int f_{1} s \anti s c_\phi d\Omega=0$,
$\int f_{2} s\anti s s_\phi d\Omega=0$,
and $\int f_{2} s\anti s c_\phi d\Omega=8\pi$.
Then, $\rho_{1,2}=\int f_{1,2} {1\over N}{dN\over d\Omega} d\Omega$.
The critical question
is with what accuracy can $\rho_{1,2}$ be determined experimentally?
The error is minimized by using projection functions which match the
angular dependence of the term of interest.  Thus,
we employ $f_1= (9/2)s\anti s s_\phi$
and $f_2=(9/2)s\anti s c_\phi$, for which $y_{1,2}=9/2$ and
$\rho_{1,2}/\delta\rho_{1,2}=
\sqrt{2/9}\rho_{1,2}\sqrt N/[1-(2/9)\rho_{1,2}^2]^{1/2}$.

We now discuss the Higgs production reactions
and the $\taum\taup$ and $t\anti t$ final state decay
modes for which the angles of Eq.~\ref{dnform} can be experimentally
determined.
Consider first the $\taum\taup$ case.  The $\tau$ decays are of two
basic types: $\tau\rta l \nu\nu$ and $\tau \rta R\nu$, where $R$
is a hadronically decaying resonance of known quantum numbers.  Together
these constitute about 95\% of the $\tau$ decays, with $BR(\tau\rta \Sigma
R\nu)\sim 58.8\%$. Thus, we employ $D=1$ and an effective branching
ratio for useful $\taum\taup$ final states of $(0.588)^2$.

In the case of $t\anti t $ decays, we employ
only the case where one top decays hadronically, and the
other leptonically, such that we are simultaneously able to determine the
exact $t\anti t$ decay axis and distinguish $t$ from $\anti t$.
Thus, we adopt an effective branching ratio for useful $t\anti t$ final
states of $2\times (2/3)\times (2/9)$ (keeping only $l=e,\mu$).
Employing {\it one} hadronic $t$ (or $\anti t$)
decay and identifying
the most energetic jet from the $\wp$ ($\wm$) with the anti-quark (quark)
leads to a depolarization factor of $D\sim 0.78$.

In order to assess our ability to experimentally measure $\rho_1$
and $\rho_2$, we have examined $\hn$ production in the reactions
$\epem\rta Z\hn$  at
a future linear $\epem$ collider and $\mupmum\rta \hn$
at a possible future $\mupmum$ collider\cite{mlc}.

For the $\epem$ collider we have adopted
the optimal energy, $\sqrt s=\mz+\sqrt 2 \mhn$,
and assumed an integrated luminosity of $85\fbi$.
For the $\mupmum$ collider we have computed the Higgs signal and the continuum
$\taum\taup$ and $t\anti t$ backgrounds assuming unpolarized beams and
a machine energy resolution
of 0.1\%, with $\sqrt s$ centered at the
value of $\mhn$.

For $\mhn$ values such that the $\epem\rta Z\hn$ production mode
is background free, the statistical
significance of a non-zero result for $\rho_1$
is that given earlier, $\nsdi=|\rho_1|/\delta\rho_1$, where
$\delta\rho_1=(9/2-\rho_1^2)^{1/2}/\sqrt{N}$, and $N$ is the number of
events {\it after including the branching ratios required to achieve the
final state of interest}: $BR_{eff}=BR(\hn\rta F\anti F)\times BR(F\anti F\rta
X)$, where the latter $F\anti F$ branching ratios to useful final $X$
states were specified above. In the case of $\rho_2$ we must actually
determine the statistical significance associated with a measurement
of $D-|\rho_2|$.  This is given by
$\nsdii=[D-|\rho_2|]/\delta\rho_2$, where
$\delta\rho_2=(9/2-\rho_2^2)^{1/2}/\sqrt N$.

In $\mupmum \rta \hn$, the continuum backgrounds must be included.
The statistical significance of a non-zero value for
$\rho_1$ is given by $\nsdi=|\rho_1|/\delta\rho_1$ with
$\delta\rho_1=[9/2-\rho_1^2+(B/S)(9/2+\rho_1^2)]^{1/2}/\sqrt S$,
where $S$ is the total number of events from $\hn$ production, and $B$
is the total number of events from the continuum background,
in the final state of interest.
For $\rho_2$ we have $\nsdii=[D-|\rho_2|]/\delta\rho_2$
with $\delta\rho_2=[9/2-\rho_2^2+(B/S)
(9/2+\rho_2^2-2\rho_2\rho_2^B)]^{1/2}/\sqrt S$,
where $\rho_2^B$ is that for the background alone.

\begin{figure}[t]  
\vspace*{-2.5cm}
\caption{The maximum statistical significances
$\nsdi$ and $\nsdii$
for $\hn\rta\taum\taup$ \solid\ and $\hn\rta t\anti t$ \dashes,
in $\epem\rta Z\hn$ ($L=85\fbi$) and $\mupmum\rta \hn$ ($L=20\fbi$)
production, after searching over all $\alpha_1$ and $\alpha_3$ values
at fixed $\mhn$ and $\tanb$.  In each case, curves for the three
$\tanb$ values of 0.5, 2, and 20 are shown. In the $\taum\taup$
($t\anti t$) mode $\nsd$ values increase (decrease) with increasing $\tanb$,
except in the case of $\mupmum\rta \hn\rta t\anti t$, where the lowest
curve is for $\tanb=0.5$, the highest curve is for $\tanb=2$,
and the middle curve is for $\tanb=20$.}
\label{asym}
\end{figure}

Our results for the maximum $\nsdi$ and $\nsdii$ values are presented
in Fig.1, where we have adopted a top quark mass of 174 GeV.
The maximum values were found by searching (holding $\tanb$ and $\mhn$ fixed)
over all values of the Higgs sector mixing angles $\alpha_1$ and
$\alpha_3$, (for the notation see Ref.~\cite{knowlesetal}).

Consider first the results for $\epem\rta Z\hn$ collisions.
{}From Fig.~1 we find that detection of CP violation
through both $\rho_1$ and $\rho_2$
is very likely to be possible for $\mhn<2\mw$ via the $\hn\rta\taum\taup$
decay mode. This is an important result given that
various theoretical prejudices suggest that the lightest
Higgs boson is quite likely to be found in this mass range.
For $\mhn$ between $2\mw$ and $2\mt$,
a statistically significant measurement of CP violation will be difficult.
For $\mhn>2\mt$, detecting CP violation in the $t\anti t$ mode
would require a somewhat larger $L$ (of order 5 times
the assumed luminosity of $L=85\fbi$ for $\tanb$ between 2 and 5).

In $\mupmum\rta\hn$ production, Fig.~1 shows that
the maximum $\nsdi$ and $\nsdii$ values
in the $\taum\taup$ mode can remain large out to large Higgs masses
if $\tanb$ is large, but that for small to moderate
$\tanb$ values the statistical significances are
better in $\epem\rta Z\hn$ collisions when $\mhn<2\mw$.
However, Fig.~1 also indicates that
the $\mupmum\rta\hn\rta\taum\taup$
channel has the advantage of possibly small sensitivity
to the $WW$ decay threshold at $\mhn\sim 2\mw$.
Such insensitivity arises when the Euler angles
$\alpha_1,\alpha_3$ are chosen so as to minimize $WW,ZZ$ couplings
(and hence $\hn\rta WW,ZZ$ branching ratios) without sacrificing
production rate. Thus, for $L=20\fbi$ $\mupmum$ collisions could allow
detection of CP violation all the way out to $2\mt$ for $\tanb\gsim 10$.
The $t\anti t$ final state extends the range of $\mhn$
for which detection of CP violation might be possible only somewhat,
and only if $\tanb$ lies in the moderate range near 2.

\vspace*{.4cm}

\noindent{\bf Acknowledgments}

BG would like to thank Davis Institute for High Energy Physics
for support during the BSMIV conference.

\end{document}